\documentclass[onecolumn]{aastex631}
\usepackage{mathtools}
\usepackage{xcolor}
\definecolor{BLUE}{rgb}{0.2,0.2,1}

\shorttitle{\sc---\,NS Core EOS\,---}
\shortauthors{\sc ---\,Cai $\&$ Li $\&$ Zhang\,---}

\graphicspath{{./}{figures/}}

\usepackage{mathrsfs}
\usepackage{amssymb}
\usepackage{amsmath}
\usepackage{graphicx}
\usepackage[normalem]{ulem}
\usepackage{bm}
\usepackage{longtable}
\usepackage{slashed}
\renewcommand\theequation{\arabic{equation}}
\usepackage[T1]{fontenc}
\usepackage{times}
\usepackage{fouriernc}

\usepackage{hyperref}
\hypersetup{colorlinks,linkcolor={cyan},citecolor={magenta},urlcolor={magenta}}

\renewcommand\sout{\bgroup \color{red} \ULdepth=-.5ex \ULset}

\renewcommand{\rm}[1]{\textrm{#1}}
\renewcommand{\d}{\mathrm{d}}

\makeatletter
\let\frontmatter@title@above=\relax
\makeatother

\let\oldbibliography\thebibliography
\renewcommand{\thebibliography}[1]{
  \oldbibliography{#1}
  \setlength{\itemsep}{1pt}
  \setlength{\baselineskip}{10.pt}
  \setlength{\lineskiplimit}{-\maxdimen}
}

\begin{document}

\title{\large Core States of Neutron Stars from Anatomizing their Scaled Structure Equations}

\author{Bao-Jun Cai\footnote{bjcai87@gmail.com}}
\affiliation{Quantum Machine Learning Laboratory, Shadow Creator Inc., Shanghai 201208, China} 

\author{Bao-An Li\footnote{Bao-An.Li$@$tamuc.edu}}
\affiliation{Department of Physics and Astronomy, Texas A$\&$M
University-Commerce, Commerce, TX 75429-3011, USA}

\author{Zhen Zhang\footnote{zhangzh275@mail.sysu.edu.cn}}
\affiliation{Sino-French Institute of Nuclear Engineering and Technology, Sun Yat-Sen
University, Zhuhai 519082, China}

\date{\today}

\fontdimen2\font=2.pt

\renewcommand{\abstractname}{\small\sc ABSTRACT}
\begin{abstract}
\small
Given an Equation of State (EOS) for neutron star (NS) matter, there is a unique mass-radius sequence characterized by a maximum mass $M_{\rm{NS}}^{\max}$ at radius $R_{\max}$. We first show analytically that the $M_{\rm{NS}}^{\max}$ and $R_{\max}$ scale linearly with two different combinations of NS central pressure $P_{\rm{c}}$ and energy density $\varepsilon_{\rm{c}}$ by dissecting perturbatively the dimensionless Tolman-Oppenheimer-Volkoff (TOV) equations governing NS internal variables. The scaling relations are then verified via 87 widely used and rather diverse phenomenological as well as 17 microscopic NS EOSs with/without considering hadron-quark phase transitions and hyperons by solving numerically the original TOV equations. 
The EOS of densest NS matter allowed before it collapses into a black hole (BH) is then obtained.
Using the universal $M_{\rm{NS}}^{\max}$ and $R_{\max}$ scalings and NICER (Neutron Star Interior Composition Explorer) and XMM-Newton mass-radius observational data for PSR J0740+6620, a very narrow constraining band on the NS central EOS is extracted directly from the data for the first time without using any specific input EOS model.
\end{abstract}

\section*{\small \textcolor{BLUE}{1. Introduction}}

Understanding the nature and Equation of State (EOS) of supradense matter in neutron stars (NSs)
\,\citep{Walecka1974,Collins1975,Chin1977,Freedman1977-1,Freedman1977-2,Freedman1977-3,
WFF1988,APR1998} has been a longstanding and shared goal of both astrophysics and nuclear physics\,
\citep{Shuryak1980,LP01,Dan02,Alford2008,LCK08,Ozel16,Watts2016,Oertel2017,Baym2018,Vidana2018,
Baiotti2019,Orsaria2019,Dexheimer2021,Lattimer2021,Lovato2022,Sorensen2023}.
In particular, novel phases and new degrees of freedom are expected to appear in NS cores where the density is the highest. Finding the EOS of densest visible matter in the Universe is an ultimate goal of astrophysics in the era of high-precision multi-messenger astronomy\,\citep{sathyaprakash2019}.
However, despite of much effort using various data and models over the last few decades information about the NS core EOS remains elusive and ambiguous\,\citep{De2018,Radice2018,Tews2018,Bauswein2019,Baym2019,Montana2019,Most2019,XL2019,XL2021,ZL2019-1,ZL2019-2,ZL2021,
Dris20,Weih2020,Zhao2020,Miao2021,Pang2021,Raaijmakers2021,Breschi2022,Ecker2022,Huth2022,Tan2022-a,Tan2022-b,Perego2022} partially because of the strong degeneracy between the core EOSs and the still poorly understood crust. 
{\it Can one extract directly the core EOS from some NS observables without using any specific model EOS?}

In this work, we show first analytically and then numerically that the maximum mass $M_{\rm{NS}}^{\max}$ and the corresponding radius $R_{\max}$ on the mass-radius sequence for any EOS scales universally with $\Gamma_{\rm{c}}\equiv\varepsilon_{\rm{c}}^{-1/2}[\widehat{P}_{\rm{c}}/(1+3\widehat{P}_{\rm{c}}^2+4\widehat{P}_{\rm{c}})]^{3/2}$ and $\nu_{\rm{c}}\equiv\varepsilon_{\rm{c}}^{-1/2}[\widehat{P}_{\rm{c}}/(1+3\widehat{P}_{\rm{c}}^2+4\widehat{P}_{\rm{c}})]^{1/2}$, respectively, where $\widehat{P}_{\rm{c}}$ is the ratio of pressure $P_{\rm{c}}$ over energy density $\varepsilon_{\rm{c}}$ at NS centers. 
The correlation between $M_{\rm{NS}}^{\max}$ and $\Gamma_{\rm{c}}$, in particular,  gives the EOS of densest matter allowed in NSs before they collapse into black holes (BHs).
Applying these universal scalings to NICER/XMM-Newton's mass-radius observational data (abbreviated as NICER since then) for PSR J0740+6620\,\citep{Fons2021,Riley2021,Miller2021,Salmi2022}, a very narrow constraining band on the NS core EOS is directly extracted without using any model EOS.

\section*{\small \textcolor{BLUE}{2. NS Radius/Mass Scalings by Perturbatively Solving the Scaled Structural Equations}}

Information about the EOS in the entire NS is normally obtained by comparing observations with predictions of a given EOS model by integrating the Tolman-Oppenheimer-Volkoff (TOV) equations\,\citep{Tolman1939,OV1939} from the NS center to its surface. 
In the following, we present a novel approach to access directly the NS core EOS without the hinderance of its still unconstrained EOSs in other regions especially the crust.

Using the energy density scale $W=G^{-1}(4\pi G\varepsilon_{\rm{c}})^{-1/2}$ and length scale $Q=(4\pi G\varepsilon_{\rm{c}})^{-1/2}$, the reduced mass $\widehat{M}\equiv M/W$, radius $\widehat{r}\equiv r/Q$, pressure $\widehat{P}\equiv P/\varepsilon_{\rm{c}}$ and energy density $\widehat{\varepsilon}\equiv \varepsilon/\varepsilon_{\rm{c}}$  (with $\varepsilon_{\rm{c}}$ being the central energy density) satisfy the dimensionless TOV equations (adopting $c=1$),
\begin{equation}\label{TOV-ds}
\frac{\d}{\d\widehat{r}}\widehat{P}=-\frac{(\widehat{P}+\widehat{\varepsilon})(\widehat{r}^3\widehat{P}+\widehat{M})}{\widehat{r}^2-2\widehat{M}\widehat{r}},~~\frac{\d}{\d\widehat{r}}\widehat{M}=\widehat{r}^2\widehat{\varepsilon}.
\end{equation}
The smallness of $\widehat{P}$ enables solving perturbatively the above coupled evolution equations of $\widehat{P}$ and $\widehat{M}$ given an EOS $P(\varepsilon)$. Starting from the NS center at $\widehat{r}=0$, the (reduced) energy density $\widehat{\varepsilon}$, pressure $\widehat{P}$ and mass $\widehat{M}$ can be expanded as $\widehat{\varepsilon}\approx1+a_1\widehat{r}+a_2\widehat{r}^2+\cdots$, $\widehat{P}\approx\widehat{P}_{\rm{c}}+b_1\widehat{r}+b_2\widehat{r}^2+\cdots$, and $\widehat{M}\approx c_1\widehat{r}+c_2\widehat{r}^2+\cdots$, respectively. Moreover, the reduced EOS $\widehat{P}(\widehat{\varepsilon})$ can be written as $\widehat{P}=\sum_{k=1}d_k\widehat{\varepsilon}^k$ with coefficients $\{d_k\}$.
The inner-relations among the coefficients $\{a_k\}$, $\{b_k\}$ and $\{c_k\}$ are determined by the NS EOS and the TOV equations.
Particularly,  as shown in the APPENDIX, all $a_k$'s and $b_k$'s with odd $k$ are zero and the first nonzero coefficient is $b_2=-6^{-1}(1+3\widehat{P}_{\rm{c}}^2+4\widehat{P}_{\rm{c}})$, 
$c_1=c_2=0$ and $c_k=a_{k-3}/k$ for $k\geq3$.

Within the perturbative approach outlined above, the reduced radius $\widehat{R}$ defined via the vanishing of pressure $\widehat{P}$ is determined to order $b_2$ by $\widehat{P}_{\rm{c}}+b_2\widehat{R}^2=0$. The resulting $\widehat{R}=[6\widehat{P}_{\rm{c}}/(1+3\widehat{P}_{\rm{c}}^2+4\widehat{P}_{\rm{c}})]^{1/2}$ multiplied by $Q\sim1/\sqrt{\varepsilon_{\rm{c}}}$ makes the physical radius $R=Q\widehat{R}$ to scale as
\begin{equation}\label{def_nuc}
R\sim
\nu_{\rm{c}}
\equiv\frac{\widehat{P}_{\rm{c}}^{1/2}}{\sqrt{\varepsilon_{\rm{c}}}}\cdot\left(\frac{1}{1+3\widehat{P}_{\rm{c}}^2+4\widehat{P}_{\rm{c}}}\right)^{1/2}.
\end{equation}
Since the reduced NS mass is $\widehat{M}(\widehat{R})\approx c_3\widehat{R}^3=\widehat{R}^3/3$, its physical mass $M_{\rm{NS}}\equiv M\sim\widehat{R}^3/\sqrt{\varepsilon_{\rm{c}}}$ then scales as 
\begin{equation}\label{def_Gammac}
M_{\rm{NS}}\sim\Gamma_{\rm{c}}\equiv\frac{\widehat{P}_{\rm{c}}^{3/2}}{\sqrt{\varepsilon_{\rm{c}}}}\cdot\left(\frac{1}{1+3\widehat{P}_{\rm{c}}^2+4\widehat{P}_{\rm{c}}}\right)^{3/2}.
\end{equation}

The above scalings relate directly the radius and mass observables of a NS with its central EOS ${P}_{\rm{c}}(\varepsilon_{\rm{c}})=P(\varepsilon_{\rm{c}})$. These scalings are general since they are direct consequences of the TOV equations themselves without assuming any particular structure, composition and EOS for the NS. Once the $\nu_{\rm{c}}$ or $\Gamma_{\rm{c}}$ is constrained within a certain range by observational data, it can be used to determine the NS central EOS.  
Moreover, the scaling for NS mass is expected to be more accurate than that for the radius which is affected more by the uncertain low-density EOS for the NS crust\,\citep{BPS71}, see FIG.\,\ref{fig_MmaxS}.
Actually the low-density EOS may affect the crust thickness and consequently the NS radius sizably while has little influence on its mass \citep{XuJ}.

Extensive studies on scaling relations among NS observables exist in the literature, see, e.g.,  Refs.\,\cite{Rhoades1974,Hartle1978,Kalogera1996,LP2005,Tsui2005,Yagi2013,Wen2019,Jiang2020} and references therein. These EOS-independent universial scalings are mostly for NS bulk properties, such as the moment of inertia ($I$), tidal Love number ($L$), (spin-induced) quadrupole moment (${Q}$), compactness ($C$), and frequencies ($f$ or $\omega$) of various oscillation modes of NSs. Knowing one or more observables, these scalings (e.g., $I$-$L$-${Q}$, $I$-$L$-$C$ or $M\omega$-$C$) enable the finding of other observables or bulk properties that have not been or hard to be measured.
They are completely different from the EOS-independent scalings between the radius and mass of most massive NSs and the combinations of the reduced core pressure and energy density that we have found and used here for the first time. Among the earlier scalings closest to ours is the one 
showing $M_{\rm{NS}}^{\max}\sim D_{\rm{M}}\varepsilon_{\rm{c}}^{-1/2}$ (see, e.g., Refs.\,\cite{Hartle1978,Kalogera1996,LP2005}) with the coefficient $D_{\rm{M}}$ estimated using either unrealistic or certain selected microscopic dense matter EOSs.
In addition, a numerical power law was given as $R\sim D_{\rm{R}}P^{\delta}$ with $\delta\approx0.23\mbox{$\sim$}0.26$\,\citep{LP01}, where $P$ is the pressure at about 1-2 times the nuclear saturation density $\rho_{\rm{sat}}$.
Since the coefficients $D_{\rm{M}}$ and $D_{\rm{R}}$ are obtained quite empirically, the corresponding predictions either have sizable EOS model dependence or lack solid/clear physical origins. 
Considering our method and most importantly, 
it is the scaling of the observables $M_{\rm{NS}}^{\max}$ and radius $R_{\max}$ with the internal variables  (different combinations of the core pressure and energy density via $\nu_{\rm{c}}$ and $\Gamma_{\rm{c}}$) that enable us to extract the relationship (core EOS) ${P}_{\rm{c}}(\varepsilon_{\rm{c}})=P(\varepsilon_{\rm{c}})$ once either one of the observables is constrained observationally, instead of just the individual value of $\varepsilon_{\rm{c}}$ or the pressure around $(1\mbox{$\sim$}2)\rho_{\rm{sat}}$ in the previous studies of scalings. Thus, as outlined in the previous section and discussed in detail in the APPENDIX, our method and results are completely novel with respect to the exisiting work in the literature.

\section*{\small  \textcolor{BLUE}{3. Validations of NS Radius/mass Scalings by Solving the Original Structural Equations}}

The maximum mass $M_{\rm{NS}}^{\max}$ and radius $R_{\max}$ characterize the densest stable NS configuration 
on the mass-radius curve before it collapses to a BH for a given NS EOS. The $M_{\rm{NS}}^{\max}$ is required to be at least as large as the currently observed most massive NS. 
Using the signals from GW170817\,\citep{Abbott2018},  the $M_{\rm{NS}}^{\max}$ (also known as the $M_{\rm{TOV}}$ for non-rotating NSs) is currently predicted to be about $2.01$\mbox{$\sim$}$2.16$ solar mass $M_{\odot}$\,\citep{Rezzolla2018}, similarly \cite{Mar2017} showed it should be $\lesssim2.17M_{\odot}$ by including the electromagnetic information). In addition, a consistent prediction of $M_{\rm{NS}}^{\max}\lesssim2.16\mbox{$\sim$}2.28M_{\odot}$ through general relativistic magnetohydrodynamics simulations was also given\,\citep{Ruiz2018}. We notice that these predictions are all somewhat EOS-model dependent.

\begin{figure}[htb!]
\centering
\includegraphics[width=8.5cm]{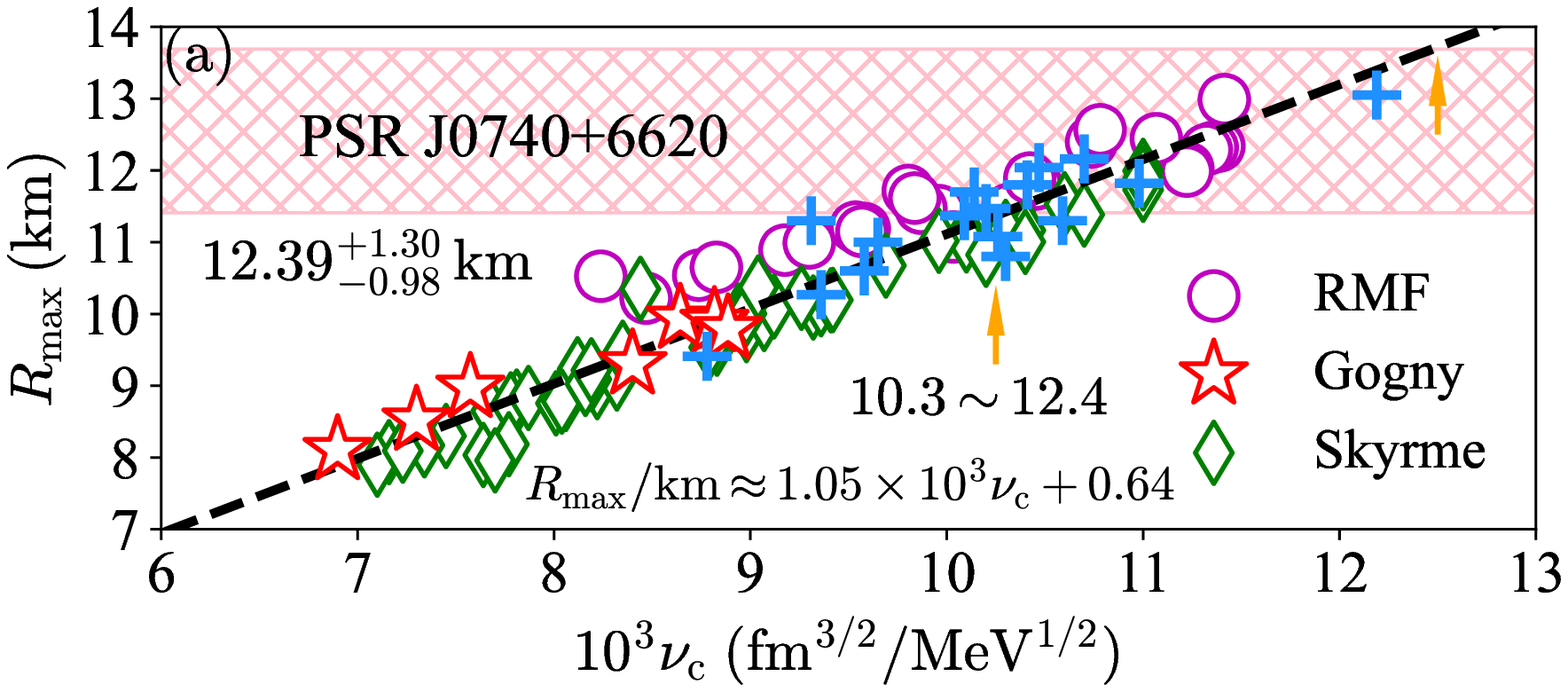}\\
\includegraphics[width=8.5cm]{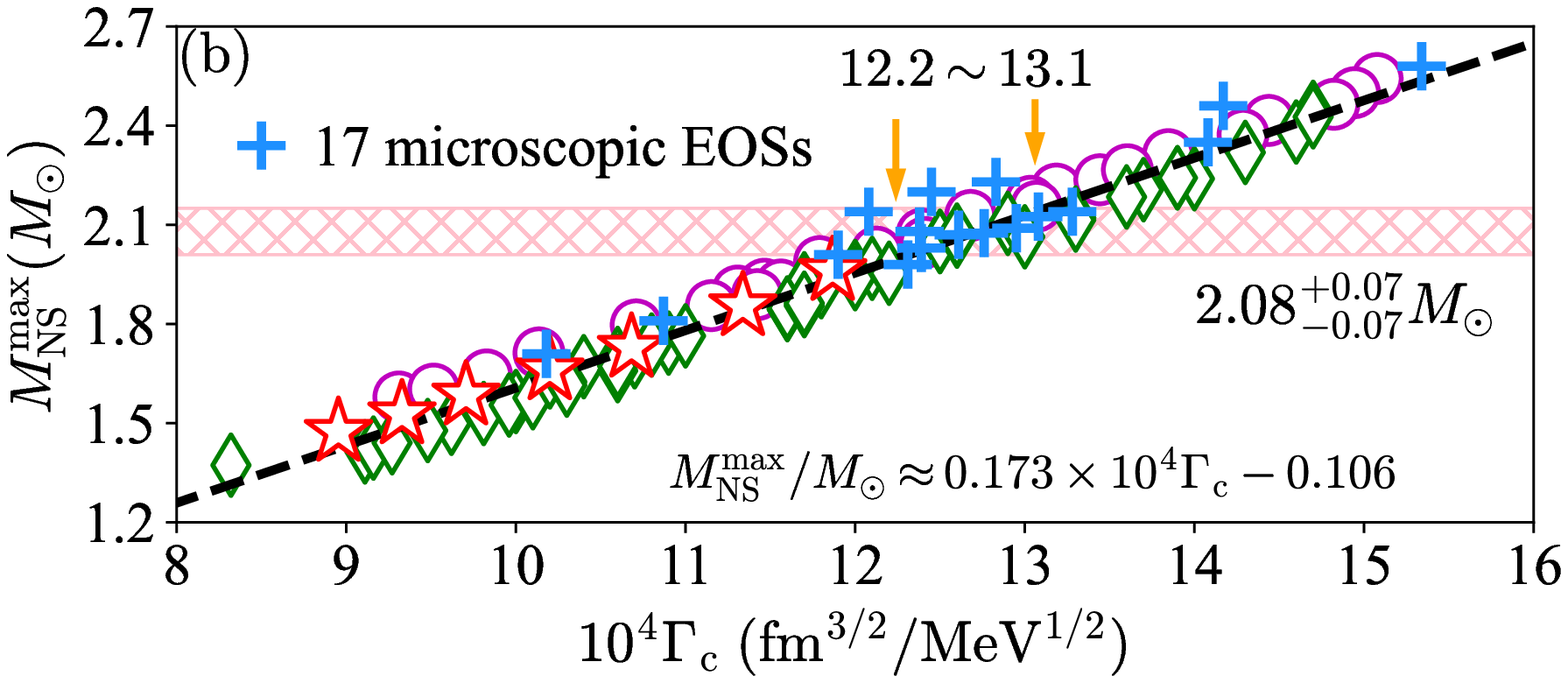}
\caption{ The $R_{\max}$-$\nu_{\rm{c}}$ (upper panel) and $M_{\rm{NS}}^{\max}$-$\Gamma_{\rm{c}}$ (lower panel) correlations with 104 EOS samples (colored symbols), 
the constraints on the mass\,\citep{Fons2021} and radius\,\citep{Riley2021} of PSR J0740+6620 are shown by the pink hatched bands. The two orange arrows and captions nearby in each panel indicate the corresponding ranges for $\nu_{\rm{c}}$ and $\Gamma_{\rm{c}}$ defined in Eq.\,(\ref{def_nuc}) and Eq.\,(\ref{def_Gammac}), respectively.
The scaling relations for $R_{\max}/\rm{km}$ and $M_{\rm{NS}}^{\max}/M_{\odot}$ are also shown (with their uncertainties given in the text).
}\label{fig_MmaxS}
\end{figure}

To test our mass and radius scalings, shown in FIG.\,\ref{fig_MmaxS} are the $R_{\max}$-$\nu_{\rm{c}}$ (upper panel) and $M_{\rm{NS}}^{\max}$-$\Gamma_{\rm{c}}$ (lower panel) correlations by using 87 phenomenological and 17 extra microscopic NS EOSs with and/or without considering hadron-quark phase transitions and hyperons by solving numerically the original TOV equations. The phenomenological EOSs 
are from three rather different classes of EOS models\,\citep{Iida1997}: the covariant relativistic mean-field (RMF) models\,\citep{Serot1986}, the non-relativistic energy density functionals (EDF) of the Gogny-like type\,\citep{cai2022nuclear} as well as the conventional Skyrme-type\,\citep{Vau1972,Stone2007,ZC16}. Within each EOS class, different EOS models are generated by perturbing the model coupling constants within their empirical uncertain ranges.  In addition, shown with the lightblue ``+'' symbols are results from using 17 extra microscopic EOS models. These include (1) the APR EOS\,\citep{APR1998} based on microscopic nuclear many-body theories using 2-body and 3-body nucleon forces; (2) the Dirac-Brueckner-Hartree-Fock method (sets MPA1\,\citep{Muther1987} and ENG\,\citep{ENG}); (3) the quark mean field model\,\citep{ALi20}; (4) the RMF calculations with hyperons\,\citep{Glen85} (as well as sets DD2\,\citep{Typel2010} and H4\,\citep{H4}); (5) the chiral mean field model with/without hyperons\,\citep{Dex08}; (6) the hybrid ALF2 EOS\,\citep{ALF2} with mixtures of nucleonic and quark matter, and (7) a few quark matter EOSs\,\citep{Prakash1995,Chu}. The observed strong linear correlations demonstrate vividly that the $R_{\max}$-$\nu_{\rm{c}}$ and $M_{\rm{NS}}^{\max}$-$\Gamma_{\rm{c}}$ scalings are nearly universal. 
It is particularly interesting to notice that EOSs allowing phase transitions and/or hyperon formations predict consistently the same scalings. 
By performing linear fits of the results obtained from the EOS samples, the scaling relations are determined to be $R_{\rm{max}}/\rm{km}\approx 1.05_{-0.03}^{+0.03}\times 10^3 \nu_{\rm{c}}+0.64_{-0.29}^{+0.29}$ with its  Pearson's coefficient about 0.958 and $
M_{\rm{NS}}^{\rm{max}}/M_{\odot} \approx 0.173_{-0.003}^{+0.003}\times 10^4 \Gamma_{\rm{c}}-0.106_{-0.035}^{+0.035}$ with its Pearson's coefficient about 0.986, respectively, here $\nu_{\rm{c}}$ and $\Gamma_{\rm{c}}$ are measured in $\rm{fm}^{3/2}/\rm{MeV}^{1/2}$ (as $\varepsilon_{\rm{c}}$ is measured in $\rm{MeV}/\rm{fm}^3$).
Moreover, the standard errors for the radius and mass fittings are about 0.39\,km and 0.05$\,M_{\odot}$ respectively for the EOS samples considered. In FIG.\,\ref{fig_MmaxS}, the condition $M^{\max}_{\rm{NS}}\gtrsim1.2M_{\odot}$ used is necessary to mitigate influences of uncertainties in modeling the crust EOS\,\citep{BPS71} for low-mass NSs. For the heavier NSs studied here, it is reassuring to see that although the above 104 EOSs predicted quite different crust properties, they all fall closely around the same scaling lines consistently, especially for the $M_{\rm{NS}}^{\max}$-$\Gamma_{\rm{c}}$ relation.

\renewcommand{\figurename}{FIG.}

Using the universal relation between $M_{\rm{NS}}^{\max}$ and $\Gamma_{\rm{c}}$ given and shown in panel (b) of FIG.\,\ref{fig_MmaxS},  the EOS of densest NS matter is found to be $P_{\rm{c}}(\varepsilon_{\rm{c}})\approx f^{2/3}\varepsilon_{\rm{c}}^{4/3}\cdot[1+4f^{2/3}\varepsilon_{\rm{c}}^{1/3}+19f^{4/3}\varepsilon_{\rm{c}}^{2/3}+\cdots]$,  which depends on the NS mass through $f\equiv (M_{\rm{NS}}^{\max}/M_{\odot}+0.106)/1730$.
It implies that a heavier NS has a smaller upper $\varepsilon_{\rm{c}}$ (as the ratio $P_{\rm{c}}/\varepsilon_{\rm{c}}$ is bounded from above) and therefore is easier to collapse into a BH\,\citep{Hawking}.

In addition, besides the correlation between $R_{\max}$ and $\nu_{\rm{c}}$ we also found a strong positive correlation between $R$ (for a given dense matter EOS) and $\nu_{\rm{c}}$ along its mass-radius (M-R) curve near the maximum-mass configuration. Considering now two points on different sides of the maximum-mass configuration with identical mass, since these two configurations would have different $\nu_{\rm{c}}$'s (as the latter positively correlates with $R$),  their $\Gamma_{\rm{c}}$'s would be essentially different, implying that the correlation between mass and $\Gamma_{\rm{c}}$ starts to break down when going away from the maximum-mass point (besides the NS crust effects mentioned earlier).
Similarly,  one needs to adopt the assumption that the NS is in (near) the maximum-mass configuration (roughly with $M_{\rm{NS}}^{\max}\gtrsim1.2M_{\odot}$) when the observational mass data is used to infer the central EOS from the mass scaling (e.g., the constraints shown in FIG.\,\ref{fig_eP-PSR740+6620}). Thus, our results are more accurate for massive NSs and the constraints on EOS essentially hold (via the factor $f\equiv (M_{\rm{NS}}^{\max}/M_{\odot}+0.106)/1730$) if stable NSs even heavier than PSR J0740+6620 are identified in the future. Moreover, since the NS maximum-mass configuration on the M-R curve is the closest point to BHs and some properties of the latter are known to be nearly universal, it is reasonable to expect that the scalings discussed in this work are most accurate for most massive NSs.

Extracting precisely the radius is much harder than the mass from observations of NSs. NICER's radius measurements for PSR J0740+6620 with its mass $\approx2.08_{-0.07}^{+0.07}M_{\odot}$\,\citep{Fons2021} have been analyzed independently using different approaches, leading to slightly different results for its radius. We report here results of our analyses using the observational radius bands extracted from three studies by the NICER Collaboration\,\citep{Riley2021,Miller2021,Salmi2022}. The observational radius band of $12.39_{-0.98}^{+1.30}\,\rm{km}$ at a 68\% confidence level from \cite{Riley2021} and the mass band given above are shown (using pink hatched bands) in the upper and lower panels of FIG.\,\ref{fig_MmaxS}, respectively. Their intersections with the two scaling lines indicate that $10.3\lesssim10^{3}\nu_{\rm{c}}/[\rm{fm}^{3/2}/\rm{MeV}^{1/2}]\lesssim12.4$ and  $12.2\lesssim{10^4\Gamma_{\rm{c}}}/[{\rm{fm}^{3/2}/\rm{MeV}^{1/2}]}
\lesssim13.1$, respectively,  in PSR J0740+6620 (indicated by the orange arrows). 
Different observational constraints on the radius may lead to different ranges for $\nu_{\rm{c}}$. We found that the radius constraint of $13.7_{-1.5}^{+2.6}\,\rm{km}$\,\citep{Miller2021} gives $11.0\lesssim10^{3}\nu_{\rm{c}}/[\rm{fm}^{3/2}/\rm{MeV}^{1/2}]\lesssim14.9$.  In addition, the very recent analysis including NICER's background estimates\,\citep{Salmi2022} found a radius of $12.90_{-0.97}^{+1.25}\,\rm{km}$, leading to $10.8\lesssim10^{3}\nu_{\rm{c}}/[\rm{fm}^{3/2}/\rm{MeV}^{1/2}]\lesssim12.9$. The core EOSs extracted using these three analyses of  PSR J0740+6620 are compared in a.Fig.\,\ref{fig_RMS} in the APPENDIX.  In the following presentation, we focus on results obtained using the radius band from \cite{Riley2021}. 

For a comparison, we notice that the hadron-quark hybrid EOS ALF2 predicts its $M_{\rm{NS}}^{\max}\approx2.09M_{\odot}$\,\citep{ALF2} that is very close to the mass of PSR J0740+6620. It also predicts $P_{\rm{c}}\approx310\,\rm{MeV}/\rm{fm}^{3}$ and $\varepsilon_{\rm{c}}\approx1100\,\rm{MeV}/\rm{fm}^3$,   therefore $\Gamma_{\rm{c}}\approx12.4\times10^{-4}\,\rm{MeV}^{1/2}/\rm{fm}^{3/2}$, which is consistent with the $\Gamma_{\rm{c}}$ extracted for PSR J0740+6620 from the mass scaling.

\section*{\small  \textcolor{BLUE}{4. Central EOS via NICER's Mass/radius Measurements and Limits for Speed of Sound}}

\begin{figure}[h!]
\centering
\includegraphics[width=8.5cm]{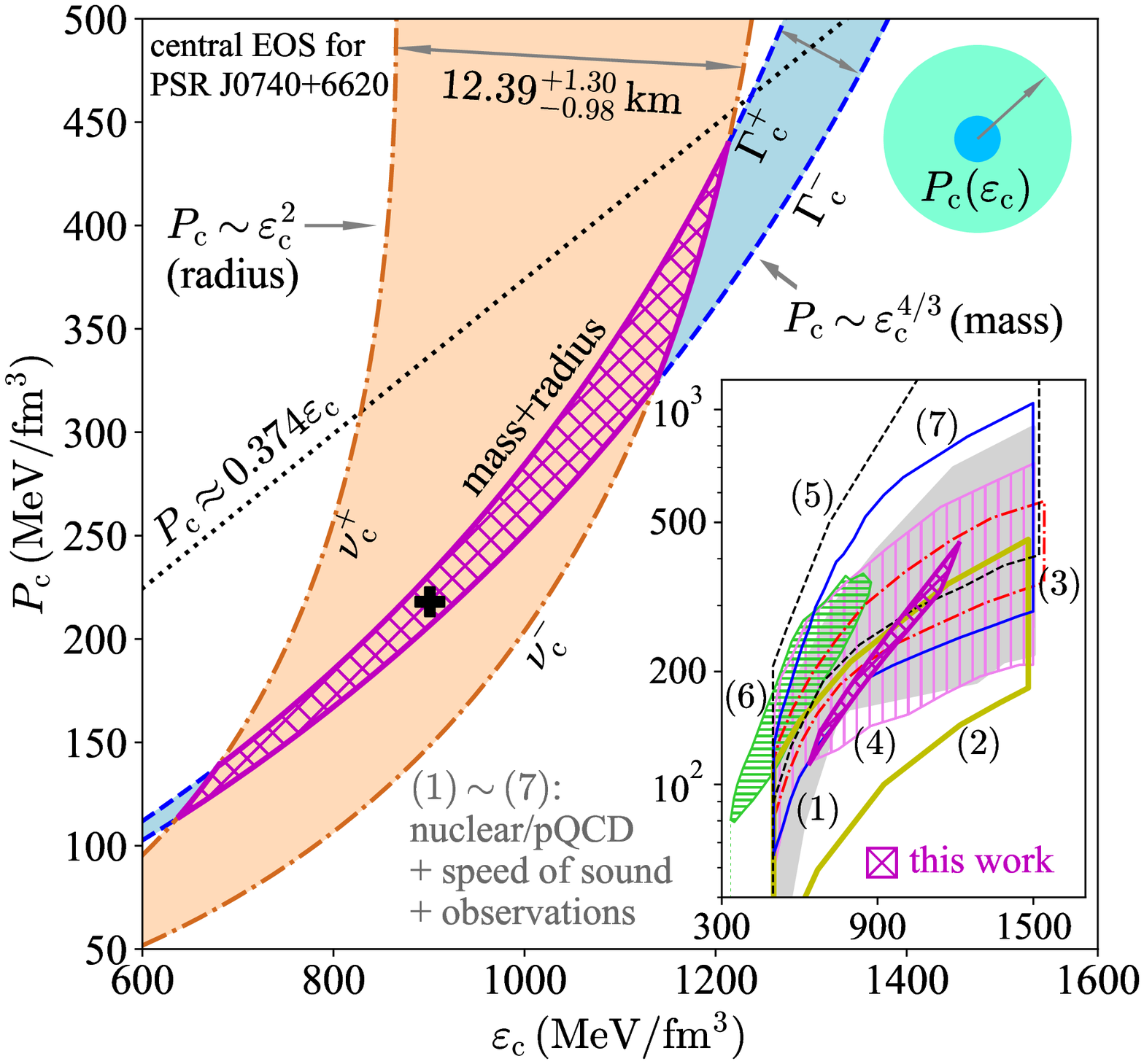}
\caption{Constraints on the central EOS $P_{\rm{c}}(\varepsilon_{\rm{c}})$ independently from the NICER measurements on the mass of $2.08_{-0.07}^{+0.07}M_{\odot}$\,\citep{Fons2021} (blue band) and the radius of $12.39_{-0.98}^{+1.30}\,\rm{km}$\,\citep{Riley2021} (chocolate band) for PSR J0740+6620, respectively.
The limit $P_{\rm{c}}/\varepsilon_{\rm{c}}\lesssim0.374$ from causality is also shown (black dotted), see Eq.\,(\ref{def_cs2}).
The inset plots our central EOS (magenta crossed band) and predictions/constraints from a few other approaches (the energy density varies from 500 to 1500$\,\rm{MeV}/\rm{fm}^3$), see text for details.}\label{fig_eP-PSR740+6620}
\end{figure}

As discussed earlier, once the $\Gamma_{\rm{c}}$ or $\nu_{\rm{c}}$ is determined within certain range by astrophysical observations, the NS central EOS $P_{\rm{c}}(\varepsilon_{\rm{c}})$ is subsequently constrained. The $\Gamma_{\rm{c}}$ and $\nu_{\rm{c}}$ extracted above from NICER's observation of PSR J0740+6620 provide two independent constraints on the NS central EOS as shown in FIG.\,\ref{fig_eP-PSR740+6620} by the blue and chocolate bands, respectively.
Since the uncertainty of the radius is currently about 3 times larger than that of the mass measurement,  the resulting constraining band of the central EOS from the radius measurement is wider. 
Consequently, the central EOS is essentially given by the mass measurement via $\Gamma_{\rm{c}}$ while the radius effectively sets the upper- and lower-limits for $\varepsilon_{\rm{c}}$. In the overlapping area of the two constraining bands,  the most probable NS central EOS can be written  as $P_{\rm{c}}(\varepsilon_{\rm{c}})\approx0.012\varepsilon_{\rm{c}}^{4/3}\cdot[1+0.047\varepsilon_{\rm{c}}^{1/3}+0.0026\varepsilon_{\rm{c}}^{2/3}+\cdots]$ where $\varepsilon_{\rm{c}}$ varies within about 640$\sim$1210$\,\rm{MeV}/\rm{fm}^3$.

To illustrate some impact of the constraining band on the NS central EOS inferred above, the inset of FIG.\,\ref{fig_eP-PSR740+6620} compares our result (shown as magenta crossed band) with predictions/constraints on the dense NS matter EOS from several other approaches: (1) (grey shaded) combination\,\citep{Ann18} of low-density nuclear,  high-density perturbative QCD (pQCD) theories and the tidal deformabilitiies from GW170817\,\citep{Abbott2018}; (2) (yellow solid) pQCD with constraint of the speed of sound at very high densities\,\citep{Kom22}; (3) (red dash-dotted) similar physical inputs as (1) combining Gaussian-process (GP) inference algorithm\,\citep{Gorda22}; (4)  (pink vertical hatched) consistent construction of EOS ensembles consistent with nuclear theories,  pQCD and astronomical observations and continuous speed of sound\,\citep{Altiparmak2022}; (5) (black dashed) nonparametric inference via GP adopting similar astronomical observations together with (6) (green horizontal hatched) the prediction on the central pressure-density of PSR J0740+6620\,\citep{Leg21}; and (7) (blue solid) the constraint from an earlier model analysis of the NICER data\,\citep{Miller2021}. Our constraining band on the central NS EOS directly from the NICER data using the mass and radius scalings without using any specific input EOS model is mostly consistent with but much narrower than the NS EOS bands from the previous studies listed above.

Observational constraints on both $\nu_{\rm{c}}$ and $\Gamma_{\rm{c}}$ enable us to determine simultaneously the NS central pressure $P_{\rm{c}}$ and energy density $\varepsilon_{\rm{c}}$.  More specifically, we obtain from the relation $\Gamma_{\rm{c}}/\nu_{\rm{c}}=\widehat{P}_{\rm{c}}/(1+3\widehat{P}_{\rm{c}}^2+4\widehat{P}_{\rm{c}})$
that $\widehat{P}_{\rm{c}}\approx0.24_{-0.07}^{+0.05}$ for PSR J0740+6620. 
Equivalently, the most probable values of the central energy density and pressure are $\varepsilon_{\rm{c}}\approx901_{-287}^{+214}\,\rm{MeV}/\rm{fm}^3$ and $P_{\rm{c}}\approx218_{-125}^{+93}\,\rm{MeV}/\rm{fm}^3$ (black ``+'' in FIG.\,\ref{fig_eP-PSR740+6620}), respectively. 
This value of central energy density is consistent with that (about 900$\sim$1200\,$\rm{MeV}/\rm{fm}^3$)
from a joint Bayesian inference using NS mass and radius constraints from GW170817 and several other astronomical observations\,\citep{Mam2021}. Using similar observational data, \cite{Brandes2023} found very recently for a $2.1M_{\odot}$ NS (with the inferred radius about 11.6\,km) that $P_{\rm{c}}\approx312_{-169}^{+226}\,\rm{MeV}/\rm{fm}^3$ and $\varepsilon_{\rm{c}}\approx904_{-327}^{+329}\,\rm{MeV}/\rm{fm}^3$. Our results are consistent with theirs but have much smaller uncertainties.

The stiffness of NS EOS can be measured effectively by the speed of sound squared $s^2\equiv\d P/\d\varepsilon$\,\citep{Bed15,Fuji22}. Using the heuristic relation $M_{\rm{NS}}\sim\widehat{R}^3/\sqrt{\varepsilon_{\rm{c}}}$ and the correlation in panel (b) of FIG.\,\ref{fig_MmaxS},  we obtain the central speed of sound squared for the maximum mass configuration,
\begin{equation}\label{def_cs2}
s_{\rm{c}}^2
=\widehat{P}_{\rm{c}}\left(1+\frac{1}{3}\frac{1+3\widehat{P}_{\rm{c}}^2+4\widehat{P}_{\rm{c}}}{1-3\widehat{P}_{\rm{c}}^2}\right).
\end{equation}
Essentially,  $s_{\rm{c}}^2$ grows (minishes) as $\widehat{P}_{\rm{c}}$ increases (decreases).
For PSR J0740+6620, using the constraints for $\widehat{P}_{\rm{c}}$ given above we obtain $s_{\rm{c}}^2\approx0.45_{-0.18}^{+0.14}$.
The condition $s_{\rm{c}}^2\leq1$ leads in addition to an upper bound $\widehat{P}_{\rm{c}}\lesssim0.374$. Therefore although the causality condition requires apparently $\widehat{P}_{\rm{c}}\leq1$, the superdense nature of core NS matter (indicated by the nonlinear dependence of $s_{\rm{c}}^2$ on $\widehat{P}_{\rm{c}}$) renders it to be much smaller.

Finally, we notice here that slightly different core EOSs are extracted by using the radii extracted by \cite{Miller2021} and \cite{Salmi2022} compared to the results presented above using the radius from \cite{Riley2021}. This is what one expects and it verifies the sensitivity of our approach. Interestingly, as shown numerically in the APPENDIX, the constraining bands on the core EOSs in PSR J0740+6620 from using the three radii\,\citep{Riley2021,Miller2021,Salmi2022} are all very narrow and they largely overlap. The remaining small widths of the constraining bands reflect mainly the observational uncertainties as no model EOS is used in extracting the NS core EOS.
As the radius uncertainty becomes smaller with the ongoing and planned new measurements using more advanced
high-precision observatories \citep{sathyaprakash2019,Watts19}, our method can be further applied to refine the core EOS of massive NSs.

\section*{\small \textcolor{BLUE}{5. Conclusion}} The mass and radius of the maximum mass configuration of any dense matter EOS scale linearly with two different combinations of their central pressure and energy density. 
The EOS of densest NS matter allowed before it collapses into a BH is obtained without using any specific model EOS.
Using NICER's mass-radius observational data for PSR J0740+6620, a very narrow constraining band on its central EOS is extracted directly from the observational data itself without using any model EOS for the first time.\\

{\it Acknowledgements: We thank Peng-Cheng Chu for helpful discussions and providing us a few quark matter EOSs.
This work is supported in part by the U.S. Department of Energy, Office of Science, under Award No. DE-SC0013702, the CUSTIPEN (China-U.S. Theory Institute for Physics with Exotic Nuclei) under
US Department of Energy Grant No. DE-SC0009971, the National Natural Science Foundation of China under Grant No.12235010.}

\vspace*{1.cm}

\appendix

\renewcommand\theequation{a\arabic{equation}}

\section*{\small A. Perturbative Treatments of the Dimensionless TOV Equations}

\indent

The dimensionless TOV equations are given in Eq.\,(\ref{TOV-ds}), where the reduced pressure $\widehat{P}$, reduced energy density $\widehat{\varepsilon}$ and the reduced mass $\widehat{M}$ are all functions of the reduced radius $\widehat{r}$. Without loosing generality, we expand them in polynomials,
\begin{align}
\widehat{\varepsilon}=&1+\sum_{k=1}a_k\widehat{r}^k\approx1+a_1\widehat{r}+a_2\widehat{r}^2+a_3\widehat{r}^3+\cdots,\\
\widehat{P}=&\widehat{P}_{\rm{c}}+\sum_{k=1}b_k\widehat{r}^k\approx\widehat{P}_{\rm{c}}+b_1\widehat{r}+b_2\widehat{r}^2+b_3\widehat{r}^3+\cdots,\\\widehat{M}=&\sum_{k=1}c_k\widehat{r}^k\approx c_1\widehat{r}+c_2\widehat{r}^2+c_3\widehat{r}^3+\cdots,
\end{align}
here $\widehat{P}_{\rm{c}}={P}_{\rm{c}}/{\varepsilon}_{\rm{c}}$ and $\widehat{\varepsilon}_{\rm{c}}=\varepsilon_{\rm{c}}/\varepsilon_{\rm{c}}=1$.
By putting these polynomials into the dimensionless TOV equations, and expanding them to the same order of $\widehat{r}$,  we obtain,
\begin{equation}
b_1=0,
\end{equation}
and
\begin{equation}
c_1=c_2=0,~~c_3=\frac{1}{3},~~c_k=\frac{a_{k-3}}{k},~~\rm{for}\;k\geq4.
\end{equation}
Moreover,
\begin{align}
b_2=&-\frac{1}{6}\left(1+3\widehat{P}_{\rm{c}}^2+4\widehat{P}_{\rm{c}}\right),\\
b_3=&-\frac{a_1}{36}\left(7+15\widehat{P}_{\rm{c}}\right),\\
b_4=&\left(-\frac{3a_2}{10}-\frac{5}{36}+\frac{1}{4}\widehat{P}_{\rm{c}}^2+\frac{1}{3}\widehat{P}_{\rm{c}}\right)\widehat{P}_{\rm{c}}-\frac{2a_2}{15}+\frac{2}{9}\widehat{P}_{\rm{c}}-\frac{a_1^2}{16}.
\end{align}

Next, the pressure could be expanded in terms of the energy density as,
\begin{equation}
\widehat{P}=\sum_{k=1}d_k\widehat{\varepsilon}^k\approx d_1\widehat{\varepsilon}+d_2\widehat{\varepsilon}^2+d_3\widehat{\varepsilon}^3+\cdots
\end{equation}
which could be rewritten as,
\begin{align}\label{ckck-1}
\widehat{P}\approx& d_1+d_2+d_3+\cdots+a_1(d_1+2d_2+3d_3+\cdots)\widehat{r}\notag\\
&+\left[(d_2+3d_3+\cdots)a_1^2+(d_1+2d_2+3d_3+\cdots)a_2 \right]\widehat{r}^2\notag\\
&+\left[\left(a_1^2d_3+2a_2d_2+6a_2d_3+\cdots\right)a_1+\left(d_1+2d_2+3d_3+\cdots\right)a_3\right]\widehat{r}^3.
\end{align}
The zeroth-order term gives the sum rule for $\{d_k\}$ as,
\begin{equation}
\widehat{P}_{\rm{c}}=d_1+d_2+d_3+\cdots=\sum_{k=1}d_k.
\end{equation}

Moreover, since $b_1=0$, we obtain
\begin{equation}
a_1(d_1+2d_2+3d_3+\cdots)=0.
\end{equation}
The speed of sound squared at the center gives another sum rule for $\{d_k\}$,
\begin{equation}
s_{\rm{c}}^2=\left.\frac{\d \widehat{P}}{\d\widehat{\varepsilon}}\right|_{\widehat{r}=0\leftrightarrow
\widehat{\varepsilon}_{\rm{c}}=1}=d_1+2d_2+3d_3+\cdots=\sum_{k=1}kd_k,
\end{equation}
which is generally nonzero, thus
\begin{equation}
a_1=0.
\end{equation}

Another argument is since $\widehat{P}$ and $\widehat{\varepsilon}$ are both even functions of $\widehat{r}$, only even terms of $\widehat{r}$ survive, one then naturally has $b_1=0$ and $a_1=0$.
According to the expression for $b_3\propto a_1$, one then obtains $b_3=0$ (and immediately $a_3=0$ could be obtained considering the coefficients in front of $\widehat{r}^3$ of Eq.\,(\ref{ckck-1})).

The next-order term is $b_4$, i.e.,
\begin{equation}\label{def_b4}
b_4=\frac{\widehat{P}_{\rm{c}}}{12}\left(1+3\widehat{P}_{\rm{c}}^2+4\widehat{P}_{\rm{c}}\right)
-\frac{a_2}{30}\left(4+9\widehat{P}_{\rm{c}}\right).
\end{equation}
Here the coefficient $a_2$ could be obtained as follows: the speed of sound squared $s^2$ is
\begin{equation}
s^2=\frac{\d\widehat{P}}{\d\widehat{\varepsilon}}=\frac{\d\widehat{P}}{\d\widehat{r}}\cdot\frac{\d\widehat{r}}{\d\widehat{\varepsilon}}=\frac{2b_2\widehat{r}+4b_4\widehat{r}^3+\cdots}{2a_2\widehat{r}+4a_4\widehat{r}^3+\cdots}=\frac{b_2+2b_4\widehat{r}^2+\cdots}{a_2+2a_4\widehat{r}^2+\cdots}.
\end{equation}
Taking it at the center gives $s_{\rm{c}}^2=b_2/a_2$, see Eq.\,(\ref{def_cs2}) for the expression of $s_{\rm{c}}^2$ (for the maximum-mass configuration).
Consequently, $a_2=b_2/s_{\rm{c}}^2$.
The $b_6$ coefficient is given by,
\begin{equation}
b_6=-\frac{1}{216}\left(1+9\widehat{P}_{\rm{c}}^2\right)\left(1+3\widehat{P}_{\rm{c}}^2+4\widehat{P}_{\rm{c}}\right)-\frac{a_2^2}{30}+\left(\frac{2}{15}\widehat{P}_{\rm{c}}^2+\frac{1}{45}\widehat{P}_{\rm{c}}-\frac{1}{54}\right)a_2-\frac{5+12\widehat{P}_{\rm{c}}}{63}a_4,
\end{equation}
here besides the coupling to $a_2$, it also couples to $a_4$ (which has no simple relation of $\widehat{P}_{\rm{c}}$ as $a_2$ does).

\renewcommand*\figurename{\small a.FIG.}
\setcounter{figure}{0}
\begin{figure}[h!]
\centering
\includegraphics[width=6.cm]{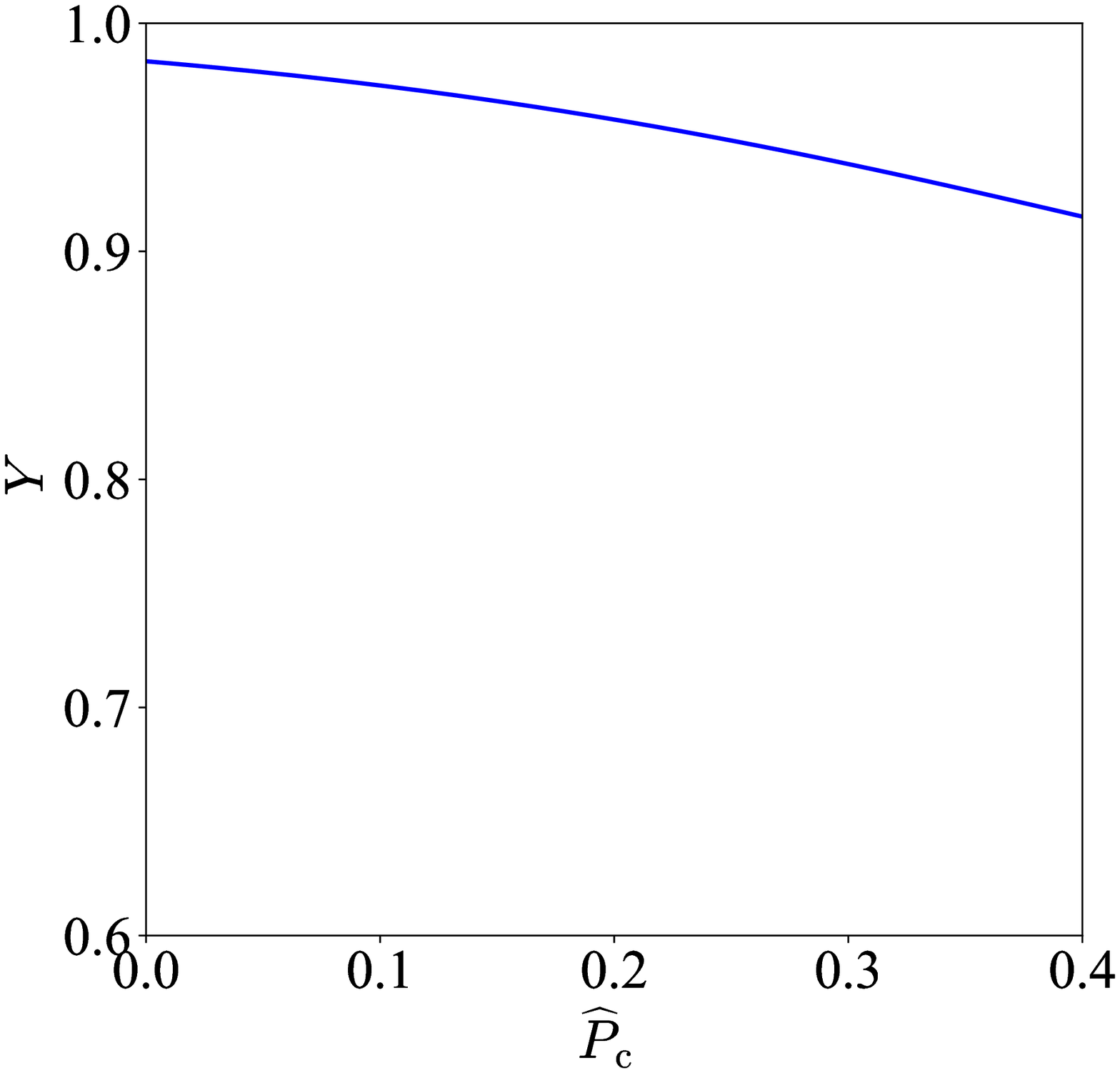}
\caption{Correction $Y$ as a function of $\widehat{P}_{\rm{c}}$.}\label{fig_Yeff}
\end{figure}

It is necessary to point out that if we truncate the expansion of the pressure to order $b_2$, then only the combination $\widehat{P}_{\rm{c}}$ comes into play since $b_2=-6^{-1}(1+3\widehat{P}_{\rm{c}}^2+4\widehat{P}_{\rm{c}})$, i.e., it is independent of the coefficients $\{a_k\}$'s.
Considering high-order contributions from like $b_4$, certain model dependence on the dense matter EOS may emerge. For instance, 
the small coefficients like $1/216,1/63$, etc., in the expression of $b_6$ reduce the effects of these high-order terms on the scaling of $\widehat{R}$ given in the main text.
As an example, we consider effects of the $b_4$-term on the dependence of $R$ versus $\nu_{\rm{c}}$. We write out
\begin{equation}
\widehat{P}_{\rm{c}}+b_2\widehat{R}^2+b_4\widehat{R}^4=0.\label{kl-1}
\end{equation}
Using the expression for $\widehat{R}^2\approx6^{-1}\widehat{P}_{\rm{c}}(1+3\widehat{P}_{\rm{c}}^2+4\widehat{P}_{\rm{c}})$ from the $b_2$-order equation, one has
\begin{equation}
\widehat{P}_{\rm{c}}+\underbrace{b_2\widehat{R}^2
+\underbrace{\left[\frac{\widehat{P}_{\rm{c}}}{12}\left(1+3\widehat{P}_{\rm{c}}^2+4\widehat{P}_{\rm{c}}\right)
-\frac{a_2}{30}\left(4+9\widehat{P}_{\rm{c}}\right)\right]}_{b_4}\underbrace{\left[\frac{1}{6} \widehat{P}_{\rm{c}}\left(1+3\widehat{P}_{\rm{c}}^2+4\widehat{P}_{\rm{c}}\right)\right]}_{\approx\widehat{R}^2}\widehat{R}^2}_{\overline{b}_2\widehat{R}^2}=0,
\end{equation}
where the effective coefficient $\overline{b}_2$ is given by,
\begin{equation}
\overline{b}_2=\underbrace{-\frac{1}{6}\left(1+3\widehat{P}_{\rm{c}}^2+4\widehat{P}_{\rm{c}}\right)}_{b_2}\times
\underbrace{\left[1+\frac{a_2}{30}\widehat{P}_{\rm{c}}\left(4+9\widehat{P}_{\rm{c}}\right)-\frac{\widehat{P}_{\rm{c}}^2}{12}\left(1+3\widehat{P}_{\rm{c}}^2+4\widehat{P}_{\rm{c}}\right)\right]}_{Y}=Yb_2,
\end{equation}
where $a_2=b_2/s_{\rm{c}}^2$ with $s_{\rm{c}}^2$ given by Eq.\,(\ref{def_cs2}) of the main text, a.FIG.\,\ref{fig_Yeff} gives the correction $Y$ as a function of $\widehat{P}_{\rm{c}}$, from which one finds that  $0.92\lesssim Y\lesssim1$ for $0\lesssim\widehat{P}_{\rm{c}}\lesssim0.4$, e.g.,  $Y\approx0.96$ for $\widehat{P}_{\rm{c}}\approx0.2$.
In this sense, the $b_4$ term does not essentially influence the scaling form for $R$ given by truncating the equation of pressure to order $b_2$.
The effects of the $b_6$-term could be estimated similarly, and the effects on the dependence of $R$ versus $\nu_{\rm{c}}$ are small.\\

\section*{\small B.  Derivations of the Core Speed of Sound $\lowercase{s}_{\rm{\lowercase{c}}}^2$}

\indent 

Using the radius scaling,
\begin{equation}
\widehat{R}\sim\left(\frac{\widehat{P}_{\rm{c}}}{1+3\widehat{P}_{\rm{c}}^2+4\widehat{P}_{\rm{c}}}\right)^{1/2},
\end{equation}
we can write the NS mass $M_{\rm{NS}}$ in the heuristic form $M_{\rm{NS}}\sim \widehat{R}^3/\sqrt{\varepsilon_{\rm{c}}}\sim F(\varepsilon_{\rm{c}})$ as,
\begin{equation}
F(\varepsilon_{\rm{c}})=\frac{\Phi}{\sqrt{\varepsilon_{\rm{c}}}}\left(\frac{\widehat{P}_{\rm{c}}}{1+3\widehat{P}_{\rm{c}}^2+4\widehat{P}_{\rm{c}}}\right)^{3/2},
\end{equation}
where $\Phi$ is a positive coefficient.
Taking derivative of $F$ with respect to $\varepsilon_{\rm{c}}$ gives,
\begin{align}
\frac{\d F}{\d \varepsilon_{\rm{c}}}
=\frac{1}{2}\frac{F(\varepsilon_{\rm{c}})}{\varepsilon_{\rm{c}}}\left[3\left(
\frac{\varepsilon_{\rm{c}}}{P_{\rm{c}}}\frac{\d{P}_{\rm{c}}}{\d\varepsilon_{\rm{c}}}-1
\right)\frac{1-3\widehat{P}_{\rm{c}}^2}{1+3\widehat{P}_{\rm{c}}^2+4\widehat{P}_{\rm{c}}}
-1\right],\label{ss-1}
\end{align}
where one uses the expression for the derivative of $\widehat{R}$ with respect to $\widehat{P}_{\rm{c}}$,
\begin{equation}
\frac{\d\widehat{R}}{\d\widehat{P}_{\rm{c}}}=\widehat{R}\left(
\frac{1-3\widehat{P}_{\rm{c}}^2}{2\widehat{P}_{\rm{c}}+8\widehat{P}_{\rm{c}}^2+6\widehat{P}_{\rm{c}}^3}
\right),
\end{equation}
and the basic relation,
 \begin{equation}
\frac{\d\widehat{P}_{\rm{c}}}{\d\varepsilon_{\rm{c}}}=\frac{1}{\varepsilon_{\rm{c}}}\left(
\frac{\d{P}_{\rm{c}}}{\d\varepsilon_{\rm{c}}}-\frac{P_{\rm{c}}}{\varepsilon_{\rm{c}}}\right).
\end{equation}
Making $\d F/\d\varepsilon_{\rm{c}}=0$ gives the speed of sound $s_{\rm{c}}^2\equiv\d P_{\rm{c}}/\d\varepsilon_{\rm{c}}$ at the center of a NS of Eq.\,(\ref{def_cs2}) in the main text. 
After obtaining the $s_{\rm{c}}^2$, we have,
\begin{equation}
\frac{\partial R}{\partial\varepsilon_{\rm{c}}}\sim
\frac{\partial}{\partial\varepsilon_{\rm{c}}}\left(\frac{\widehat{R}}{\sqrt{\varepsilon_{\rm{c}}}}\right)
=\left(\frac{s_{\rm{c}}^2}{\widehat{P}_{\rm{c}}}-1\right)\frac{1-3\widehat{P}_{\rm{c}}^2}{1+3\widehat{P}_{\rm{c}}^2+4\widehat{P}_{\rm{c}}}-1=-\frac{2}{3},
\end{equation}
i.e., as $\varepsilon_{\rm{c}}$ increases, the radius $R$ decreases, as expected.\\

\section*{\small C.  Analyses using Different Radius Constraints on PSR J0740+6620}

\indent

\begin{figure}[h!]
\centering
\includegraphics[width=7.5cm]{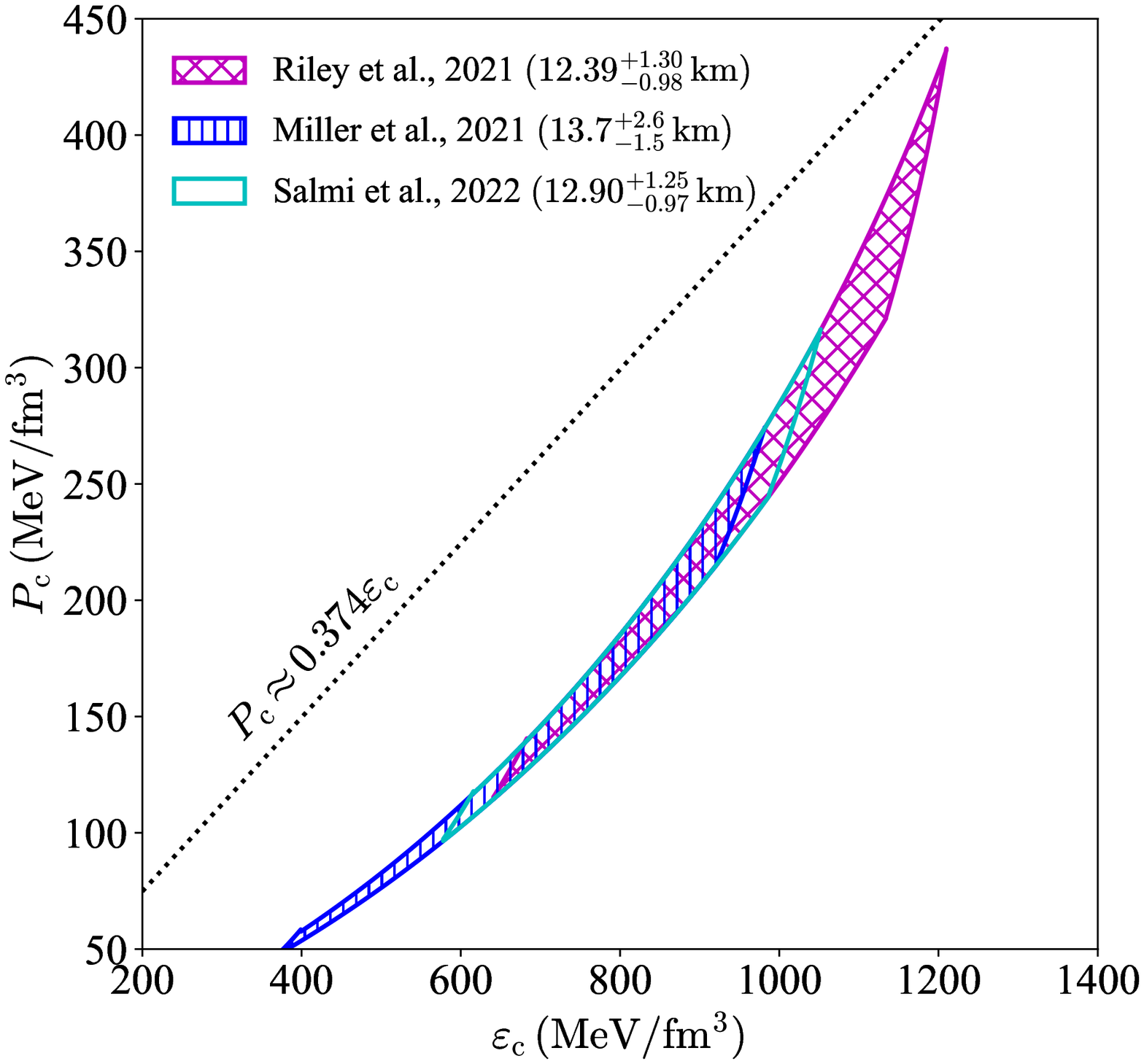}
\caption{Constraining bands for the central EOS of PSR J0740+6620 using different radius constraints\,\citep{Riley2021,Miller2021,Salmi2022}.}\label{fig_RMS}
\end{figure}

In Table \ref{sstab}, we show the radius constraints on PSR J0740+6620 using different inference/analysis methods\,\citep{Riley2021,Miller2021,Salmi2022}, as well as the resulted factor $\nu_{\rm{c}}$, central energy density $\varepsilon_{\rm{c}}$, central pressure $P_{\rm{c}}$ and the central speed of sound squared $s_{\rm{c}}^2$.
The central EOS is shown in a.FIG.\,\ref{fig_RMS}, from which one can find that the constraining bands for $P_{\rm{c}}(\varepsilon_{\rm{c}})$ in all three cases are narrow. The uncertainties on $P_{\rm{c}}(\varepsilon_{\rm{c}})$ mainly reflect the observational uncertainty on the radius of PSR J0740+6620.

\renewcommand{\arraystretch}{1.35}
\begin{table}[tbh!]
\hspace*{-2.cm}
\centering{
\begin{tabular}{|c|c|c|c|c|c|} 
  \hline
Reference& radius\,(km) &$10^3\nu_{\rm{c}}$&$\varepsilon_{\rm{c}}$&$P_{\rm{c}}$&$s_{\rm{c}}^2$\\\hline\hline
       \cite{Riley2021} &$12.39_{-0.98}^{+1.30}$&$11.2_{-0.9}^{+1.2}$&$902_{-287}^{+214}$&$218_{-125}^{+93}$&$0.45_{-0.18}^{+0.14}$\\\hline
    \cite{Miller2021}  &$13.7_{-1.5}^{+2.6}$&$12.4_{-1.4}^{+2.5}$&$656_{-339}^{+187}$&$124_{-99}^{+53}$&$0.32_{-0.14}^{+0.08}$\\\hline
    \cite{Salmi2022} &$12.90_{-0.97}^{+1.25}$&$11.7_{-0.9}^{+1.2}$&$794_{-235}^{+181}$&$173_{-89}^{+69}$&$0.39_{-0.13}^{+0.09}$\\\hline
    \end{tabular}}
\caption{Most probable values for the central energy density $\varepsilon_{\rm{c}}$, central pressure $P_{\rm{c}}$ and the central speed of sound squared $s_{\rm{c}}^2$ for PSR J0740+6620, here its mass about $2.08_{-0.07}^{+0.07}M_{\odot}$ and the corresponding $12.2\lesssim{10^4\Gamma_{\rm{c}}}/[{\rm{fm}^{3/2}/\rm{MeV}^{1/2}]}
\lesssim13.1$ are used for all three cases.
Here $10^3\nu_{\rm{c}}$ is measured in unit $\rm{fm}^{3/2}/\rm{MeV}^{1/2}$, $\varepsilon_{\rm{c}}$ and $P_{\rm{c}}$ are measured in $\rm{MeV}/\rm{fm}^3$.
}
\label{sstab}
\end{table}

\newpage

\end{document}